# Deep Learning Adapted Acceleration for Limited-view Photoacoustic Computed Tomography

Hengrong Lan, Jiali Gong, and Fei Gao, *Member, IEEE*

*Abstract*—Photoacoustic imaging (PAI) is a non-invasive imaging modality that detects the ultrasound signal generated from tissue with light excitation. Photoacoustic computed tomography (PACT) uses unfocused large-area light to illuminate the target with ultrasound transducer array for PA signal detection. Limited-view issue could cause a low-quality image in PACT due to the limitation of geometric condition. The model-based method is used to resolve this problem, which contains different regularization. To adapt fast and high-quality reconstruction of limited-view PA data, in this paper, a model-based method that combines the mathematical variational model with deep learning is proposed to speed up and regularize the unrolled procedure of reconstruction. A deep neural network is designed to adapt the step of the gradient updated term of data consistency in the gradient descent procedure, which can obtain a high-quality PA image only with a few iterations. Note that all parameters and priors are automatically learned during the offline training stage. In experiments, we show that this method outperforms the other methods with half-view (180 degrees) simulation and real data. The comparison of different model-based methods show that our proposed scheme has superior performances (over 0.05 for SSIM) with same iteration (3 times) steps. Furthermore, an unseen data is used to validate the generalization of different methods. Finally, we find that our method obtains superior results (0.94 value of SSIM for in vivo) with a high robustness and accelerated reconstruction.

## I. INTRODUCTION

IN computational imaging, a high-quality image helps us obtain effective information with high resolution. However, limited hardware and circumstances (e.g., motion disturbance, limited angle, noise) could decrease the quality of imaging result. The ill-posed inverse problem is formed if we cannot capture enough valid data. Therefore, many researchers contributed novel algorithms to obtain high-quality images in different ill-posed conditions.

Photoacoustic computed tomography (PACT) is an emerging imaging modality [1, 2]. PAI can detect endogenous chromophores or exogenous contrast agents based on PA effect [3], which refers to ultrasound generation by short-pulsed laser excitation. In PACT, unfocused light is used to excite PA signals, and many detectors (transducer array) are placed at different positions for signal detection. PACT has been applied in small animal imaging, and breast cancer diagnosis [4-7].

In general, we should acquire enough channels' data at different positions to reconstruct a high-quality image. However, the ill-conditions also exist in PACT, e.g. limited detection angle or limited number of detector. To address it, some algorithms were proposed to solve this ill-posed inverse problem by the model-based schemes (variational methods) [8, 9]. These methods unrolled iterative model-based and iterate several times. Generally speaking, this approach reconstructs the PA image by solving an optimization problem.

Convolutional neural networks (CNN), with rapid development of deep learning (DL), have been actively studied for different areas. Recently, deep learning are gaining attention for solving inverse problem in PACT [10, 11]. One of DL schemes is non-iteratively learning the relationship of different data with a greater mapping capability. For example, raw data enhancements include channel data interpolation, bandwidth recovery, and noise cancellation [12, 13]; direct mapping from signals to image [14-17]; low-quality PA image inpainting includes artifacts removal, resolution enhancement [18-20]. However, these methods have increased overfitting risks. An alternative method is an iterative scheme that combines DL with an unrolled iterative model-based image reconstruction algorithm. This approach efficiently finds the solution with learning parts of the optimization procedure [21-23]. Alternatively, a learned regularizer/denoiser is designed to penalize the data consistency from training datasets [24, 25]. However, the above procedure requires many steps of iterations,

Hengrong Lan and Jiali Gong contributed equally to this work.
Hengrong Lan is with the Hybrid Imaging System Laboratory, Shanghai Engineering Research Center of Intelligent Vision and Imaging, School of Information Science and Technology, ShanghaiTech University, Shanghai 201210, China, with Chinese Academy of Sciences, Shanghai Institute of Microsystem and Information Technology, Shanghai 200050, China, and also with University of Chinese Academy of Sciences, Beijing 100049, China (e-mail: lanhr@shanghaitech.edu.cn).

Jiali Gong is with the Hybrid Imaging System Laboratory, Shanghai Engineering Research Center of Intelligent Vision and Imaging, School of Information Science and Technology, ShanghaiTech University, Shanghai 201210, China (e-mail: gongjl@shanghaitech.edu.cn).

Fei Gao is with the Hybrid Imaging System Laboratory, Shanghai Engineering Research Center of Intelligent Vision and Imaging, School of Information Science and Technology, ShanghaiTech University, Shanghai 201210, China (*e-mail: gaofei@shanghaitech.edu.cn).

and the total computational cost scale with the dimensions of signals and image. Furthermore, different methods require tuning initial parameters for different data lacking a sophisticated step-size control, which could influence convergence and acceleration. Therefore, the major issue of this scheme is how to achieve fast convergence.

Namely, we expect that the model-based approach can significantly accelerate imaging rate with high quality. In this work, we adopt the second scheme and further improve the performance. We introduce an efficient adapted gradient form that combines DL and gradient descent with an adaptive regularization for accelerated PACT reconstruction, which we call deep adapted variation (DAV). A CNN embeds the conventional model-based formulation as a variational form. Meanwhile, the CNN constrains the regularized information from the off-line training procedure without the handcrafted prior as the gradient form. After training, an adaptive step-size is embedded in each iterative reconstruction that we can automatically obtain the high-quality image. Since we do not change the data consistency term in DAV, it could guarantee the generalization of our method. Furthermore, DAV can automatically learn the proportion of each iteration without the tuning parameters. To evaluate the performance, we demonstrate different methods on simulation and experimental data. We train the DAV on different limited-view datasets (180° view). In addition, an unseen data is used to compare the generalization of different methods, which is different from the training data. These experimental results show that our method outperforms other previous methods with a robust generalization.

Our contributions can therefore be summarized as follows:
- We propose an iterative form, named as DAV, which combines DL and gradient descent. For DAV, parts of components in model-based method including regularization, free parameters tuning, and adaptive step are learnt from training dataset. Furthermore, we only process the current result, and do not change the data consistency term, which guarantees a robust performance. Specifically, DAV boost the convergence rate and accelerate the iteration speed.
- To validate our method, we evaluate DAV and other DL-based methods on simulated vessel data and a public *in-vivo* mice data. The results show that our method significantly improves reconstruction quality compared to conventional model-based method with only 3 iterations. Meanwhile, our method has a better performance compared with other DL-based methods at same iterations.
- In addition, we release all codes for this work, which can be found at: https://github.com/chenyilan/DAV .

## II. METHOD

### A. Photoacoustic Computed Tomography

For PACT, a nanosecond pulsed laser illuminates the tissue, and chromophores could absorb the photons. The energy will convert to heat, and induce the initial pressure change with the optical fluence $F$:

$$p_0 = \Gamma_0 \eta_{th} \mu_a F, \quad (1)$$

where $\Gamma_0$ is the Gruneisen coefficient, $\eta_{th}$ is the conversion efficiency, and $\mu_a$ is the optical absorption coefficient. The pressure will propagate in the medium as a spatiotemporal equation that:

$$\left(\nabla^2 - \frac{1}{s^2}\frac{\partial^2}{\partial t^2}\right) p(\mathbf{r},t) = -\frac{\beta}{C_P}\frac{\partial H(r,t)}{\partial t}, \quad (2)$$

where $s$ is the speed of ultrasound, H is the heating function, $\beta$ and $C_P$ denote the thermoelastic expansion and heat parameters. For different positions $\mathbf{r}'$ of detectors, we derive the pressure field by Green function:

$$p(\mathbf{r},t) = \frac{1}{4\pi s^2}\frac{\partial}{\partial t}\left[\frac{1}{st}\int d\mathbf{r}' p_0(\mathbf{r}')\delta\left(t - \frac{|\mathbf{r}-\mathbf{r}'|}{s}\right)\right]. \quad (3)$$

Note that the detectors measure these signals at the boundary and constitute the PA signal $b$ from $t=0$ to $t=T$. Furthermore, the sensing procedure could contain the transfer relation of ultrasound system. We can model the procedure from $p(\mathbf{r}, t)$ to $b$ as a linear operator $\mathcal{A}$:

$$b = Ap. \quad (4)$$

Eq. (4) describes the forward procedure ($A$ is forward operator) that convers initial pressure $p$ to measured time series data $b$. Similarly, the corresponding image reconstruction from $b$ to $p$ is called the inverse problem.

### B. PA image reconstruction

An intuitive reconstruction is computing the inverse of $\mathcal{A}$ with the matrix form. However, for most PA reconstruction problems, the $\mathcal{A}$ is intractable to compute as matrix.

The reconstruction can be classified as direct scheme and model-based scheme. Likewise, the development of DL-based reconstruction approaches still follow the paths of these two schemes as mentioned above. Direct scheme is especially attractive in practice as they exhibit simple implements. We can directly compute the inversion of Eq. (3) using universal back-projection (UPB) [26]:

$$p_0(\mathbf{r}) = \frac{1}{\Omega_0}\int_{S_0}\left[2p(\mathbf{r}_0,t) - \frac{2t\partial p(\mathbf{r}_0,t)}{\partial t}\right]\frac{\cos\theta_0}{|\mathbf{r}-\mathbf{r_0}|^2}dS_0, \quad (5)$$

where $\theta_0$ is the angle between the vector pointing to the reconstruction point $\mathbf{r}$ and transducer surface $S_0$. Meanwhile, the adjoint [27] and time-reversal [28] approaches are computationally efficient in different setting by solving a single wave equation. Nevertheless, they are inadequate for dealing with the data in limited conditions, e.g., limited-view, sparse sampling.

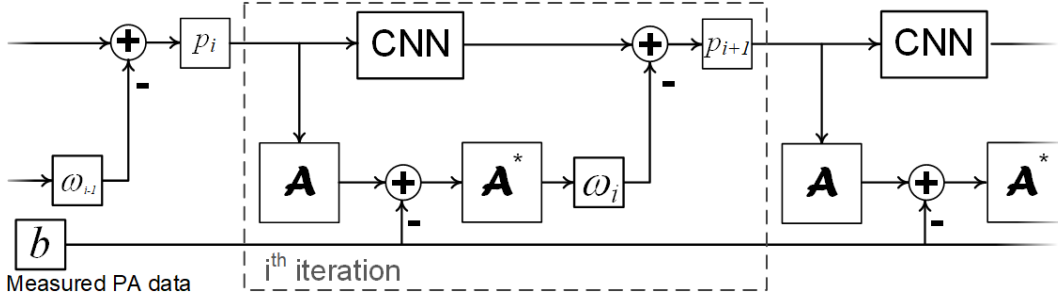

Fig. 1. The procedure of unfolded DAV. $p_i$ is the current estimated image, and ω is a learnable parameter.

The model-based approach is used to recover the initial pressure $p$ from the sub-sampled or limited-view PA data $b$ as a minimization problem:

$$p^* = \arg\min_p \frac{1}{2}\|Ap - b\|_2^2 + \alpha R(p), \quad (6)$$

where $\|Ap - b\|_2^2$ is data consistency term, $R(p)$ indicates the regularization, which constraints the prior information about $p$. $\alpha$ is the regularization parameter to tune the proportion of the data consistency term and the regularization term. An effective regularization can improve the quality of image and reconstruct the PA image with less measured channels, e. g., L1 , Tikhonov regularization , Total variation (TV) [29]. However, the suitable regularization is difficult to choose for different setting. For Eq. (6), we can employ a gradient descent method for a differentiable $R(p)$ by computing the iteration:

$$p_{i+1} = p_i - \lambda \left[ A^*(Ap_i - b) + \alpha \frac{\partial R(p_i)}{\partial p} \right], \quad (7)$$

where $\lambda$ is the step-size of each iteration, $A^*$ is the ajoint operator of $A$. $A^*(Ap_i - b)$ is the gradient of the data consistency. However, some regularizations are not differentiable, such as L1 and TV. For these regularizations, the proximal gradient method is used:

$$p_{i+1} = \text{prox}_{R,\alpha}\left(p^i - \lambda A^*(Ap_i - b)\right), \quad (8)$$

where $\text{prox}_{R,\alpha}$ is the proximal operator. We have to find different proximal operator for different $R$, and converges a final image with a large number of iterations.

### C. Deep adapted variation

Different regularizations take different functions that you want to use to constrain the structure of image. For instance, L1 is designed to perfectly recover the image with a sparse structure, which is usually applied for compressed sensing applications. However, most images contain more involved information, which causes that different prior could be constrained with multi-regularizations. More priors cause more parameters for different terms, and further influence the speed of convergence. A simple $R$ term (e. g., L1, TV) cannot lead to optimal results in this case, prior structure should be further analyzed. In addition, for a new regularizer, a mathematical analysis of differentiability is necessary, which determines whether the gradient descent can be used.

Recently, learnable regularizations have been demonstrated to constrain the PA image from a large number of dataset. As an example in [22], the authors learned a proximal operator based on the structure of Eq. (8), which could implicitly learn a prior during training. This scheme can form a validated prior for similar data, and the results could be over-fitted to single domain since the data consistency is encoded by model. Furthermore, a CNN-based regularization was presented for compressed sensing PACT.

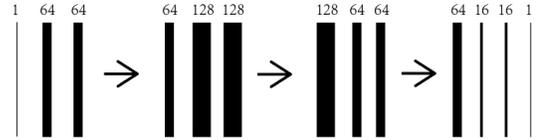

Fig. 2. The structure of CNN. Each line means a 3×3 convolutional layer followed by a BatchNorm2d layer and a leaky_relu layer, except that the last convention is 1×1 without batch normalization and leaky_relu. The number on the top means the number of channels.

In contrast to [22], we propose the DAV form to learn a variation of regularization using a CNN model in this work. Instead of learning an operator by feeding the current $p_i$ and the gradient of $(Ap_i - b)$, we propose to learn a prior step with a primary gradient decent in Eq. (7) for each iteration:

$$p_{i+1} = CNN(p_i) - \omega_i(A^*(Ap_i - b)), \quad (9)$$

where $\omega_i$ is a learnable step-size for each iteration. Note that we train our model layer by layer, and the step-size could be changed at different iteration. Namely, the step-size is adaptive. The procedure of unfolded DAV is illustrated in Fig. 1, where $p_i$ is computed by the gradient of data consistency and the output of CNN for each iteration, respectively. Note that the data consistency adds to the output image of CNN through a learnable step ω. This implies that, without data consistency, the CNN model only learns the prior information about current $p_i$ as an image processing step. In this way, we only extract the features of $p_i$ with adaptive step-size, and the $R$ is learned implicitly from the data. This procedure enhances the generalization so that DAV can handle the unseen data since we retain the data consistency term to lead the direction of convergence.

Furthermore, the architecture of the CNN to implement Eq. (9) is illustrated in Fig. 2. In particular, the network only processes the gradient and prior structure about current image instead of processing gradient of image.

In training procedure, we should pre-define a maximum iteration $i_{max}$, and we minimize the difference between ground-truth image $p_{gt}$ and the iterated image of each iteration $p_i$ using MSE function. Namely, our loss function of each iteration is given by:

$$\text{Loss}(x_i) = \|p_i - p_{gt}\|_2^2. \quad (10)$$

Since the operators $A$ and $A^*$ need to be evaluated for each iteration, we choose a layer-by-layer training scheme. In our work, for all learned iterative methods, $i_{max}$ is set as 3. Finally, we summarize the layer-by-layer training procedure in **Algorithm 1**. Once the training procedure is accomplished, the evaluation of DAV is equivalent to Algorithm 1 without *TrainIter*.

---

**Algorithm 1** The training procedure

---

1: **Input:** $p_1 = A^*b$, $i_{max}=3$.
2: **for** $i=1$ to $i_{max}$ **do**
3:     Compute $A^*(Ap_i-b)$;
4:     **function** *TrainIter* ($p_i$, $p_{gt}$, $A^*(Ap_i-b)$)
5:         Training whole dataset and updating $\Theta$ with given epochs;
6:     **end function**
7:     $p_{i+1} \leftarrow \text{CNN}(p_i) + A^*(Ap_i-b)$;
8:     $i \leftarrow i+1$;
9: **end for**

---

## III. EXPERIMENTS

To test and verify the performance of DAV, clinical data is most persuasive. However, clinical data of PACT, as a newly developed imaging technology, is so rare. Existing data sets usually do not include enough data to train a well-performing network. As a result, we usually have to implement data-driven works with simulated data or animal *in-vivo* data. As comparison study, we consider three different strategies to compare with DAV: 1) total variation, 2) a U-Net [30] as post-processing, 3) Deep Gradient Descent (DGD [22]), 4) Recurrent Inference Machines (RIM) [23]. These methods will be validated on same dataset and compared with our proposed method.

### A. Simulated data

We first implemented our experiment on simulated data. We set up a simulation environment in MATLAB toolbox k-Wave [31] to perform PA simulation. The whole simulated region is 40mm×40mm divided into 380 × 380 grids. The region of interest (ROI) is of 26.95mm × 26.95mm size with the 64 transducers at the top. The radius of half-ring-shape transducer is 19 mm. The center frequency of the transducer is 2.5 MHz with 80% fractional bandwidth. The speed of sound is set as 1500 m/s. Final images should be cut from the reconstructed image, including 256 × 256 pixels.

We chose a public dataset, DRIVE [32], to deploy with initial pressure distribution. Some operations are used to expand the data volume: 1). the complete blood vessel of fundus oculi is factitiously segmented into four equal parts; 2). randomly rotational transform and superpose two fragments. Finally, these data are loaded into k-Wave simulation toolbox as the initial pressure distribution. The training set consists of 3600 images and the test set consists of 400 images.

### B. In-vivo data

After achieving a good performance in the experiment on the simulated data, the model should be applied to *in-vivo* data experiment, which is more complicated and practical. The animal *in-vivo* data is from the public data set MOST-Abdomen [17], which is generated by PA scan of mice's abdomen. Due to the different demand of reconstruction setting, we rebuild the ground-truth images. The transducer contains 86 channels' elements with a 180° view. The center frequency of the transducer is 5 MHz. Final images have 128 × 128 pixels. There are 629 images in the training set and 70 images in the test set.

### C. Implementation

For the DAV, we use Pytorch [33] to train our CNN and obtain the predicted result. When we get the result of current iteration $p_i$, the result is read by MATLAB to precompute gradient $A^*(Ap_i\text{-}b)$ for the next iteration with the help of k-Wave toolbox. Then the new iteration starts. The CNN for the new iteration will be trained and tested. The total number of iterative steps is 3.

In each iteration, we should mix the training set and test set for cross validation. For comparing different methods, the selection is recorded and should not change from test set.

We chose Adam optimizer [34] when training, and the number of epochs is 300. The learning rate is around 0.0004, and batch size is 32. On simulated data set, the learning process of CNN roughly costs 10 hours on GPU (GeForce TITAN RTX). Besides, the calculation of gradient on MATLAB costs around 4 hours. On *in-vivo* dataset, due to the small number of data, it takes half an hour to train CNN and 2 hours to calculate gradient.

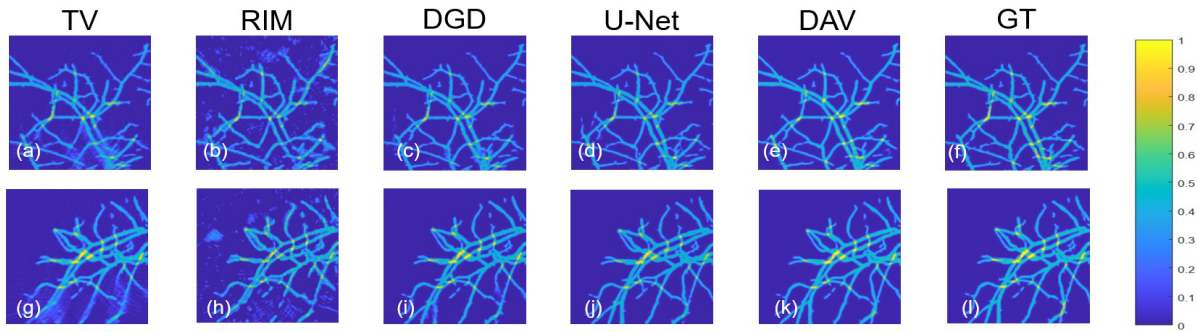

Fig. 3. The samples of simulated results. (a)-(f) are sample 1; (g)-(l) are sample 2. (a) (g) total variation with 30 iterations; (b) (h) recurrent inference machines; (c) (i) deep gradient descent; (d) (j) non-iterative U-Net; (e) (k) our method; (f) (l) ground-truth.

## IV. RESULTS

The quantitative index of evaluating the performance of various reconstruction methods is Peak Signal to Noise (PSNR) and Structural Similarity (SSIM), which show the similarities between original images and reconstructed images. TV is the typical numerical regularization item, usually representing the baseline of PACT reconstruction problem. U-Net has an outstanding performance in the field of image semantic segmentation, and is widely used in medical imaging community. Note that RIM is different from other methods in that it has two hidden variable $st_1$ and $st_2$, which is also recorded in each iteration and read by Pytorch in the next iteration.

TABLE I
THE QUANTITATIVE RESULTS OF SIMULATED DATA WITH DIFFERENT ITERATIONS (MEAN±STD)

|   | DAV | | DGD | | RIM | |
|---|---|---|---|---|---|---|
|   | SSIM | PSNR | SSIM | PSNR | SSIM | PSNR |
| 1 | 0.8906±0.0250 | 26.1486±2.0107 | 0.8783±0.0249 | 26.1122±1.9711 | 0.6314±0.0374 | 21.6592±1.7684 |
| 2 | 0.9451±0.0167 | 29.1130±2.0182 | 0.9409±0.0187 | 28.8580±2.5922 | 0.6265±0.0426 | 21.3911±1.2304 |
| 3 | **0.9676±0.0102** | 31.6155±1.7476 | 0.9642±0.0131 | **32.0170±3.1718** | 0.6857±0.0396 | 22.8294±1.4250 |

### A. Simulated data

In simulated experiments, the average PSNR and SSIM of whole test results are listed in TABLE I. For the result of U-Net, the quantitative performance achieved $0.9698 \pm 0.0107$ (SSIM) and $31.8353 \pm 3.0256$ (PSNR). For TV reconstruction, the leaning rate $\lambda$ is set as 2 and the coefficient $\alpha$ is set as 0.04. The number of iterations is 20. Two examples are shown in Fig. 3. The TV result has some blur at the end of the blood vessel. RIM result relatively has the most noise. The other three methods can generate clear enough images to recognize the blood vessels despite having some detail artifacts.

### B. in-vivo data

The simulated test proved that our method has a good performance on simulated data. We also tested our method on *in-vivo* data. Three examples are shown in Fig. 4. In contrast to iterative methods, U-Net reconstruction is more unstable. Its performance changes drastically according to the composition of training set and test set. Moreover, U-Net has a large number of parameters, thus when training set is relatively small, such as the *in-vivo* data set, it is likely to be over-fitted.

The quantitative evaluations of test data are listed in TABLE II. The PSNR and SSIM results of U-Net reconstructed images are $22.8534 \pm 4.1956$ and $0.9238 \pm 0.0169$ respectively. For TV reconstruction, the leaning rate $\lambda$ is set as 0.0015 and the coefficient $\alpha$ is set as 0.0002. On account of the small gap between initial images and ground truth images, small leaning rate and coefficient can help increase stability. After 30 iterations, the PSNR and SSIM results of TV reconstructed images are $23.7190 \pm 3.3416$ and $0.8631 \pm 0.0261$, respectively. From Fig. 4, the TV and RIM results cannot clearly show the subtle color changes at the contours of the object. DGD results are much clearer, but still have some blur at the contours. U-net results has the least background noise, but some artifacts make the results not as similar to the ground-truth as our results.

TABLE II
THE QUANTITATIVE RESULTS OF IN-VIVO DATA (MEAN±STD)

|   | DAV | | DGD | | RIM | |
|---|---|---|---|---|---|---|
|   | SSIM | PSNR | SSIM | PSNR | SSIM | PSNR |
| 1 | 0.9043±0.0212 | 29.8877±2.5567 | 0.8648±0.0211 | 26.2491±2.8282 | 0.8626±0.0288 | 23.1232±2.5472 |
| 2 | 0.9275±0.0145 | 33.1818±2.3627 | 0.8770±0.0173 | 28.1606±3.1028 | 0.8727±0.0256 | 25.3953±3.4914 |
| 3 | **0.9413±0.0191** | **35.0995±2.4308** | 0.8846±0.0279 | 28.2611±3.1837 | 0.8678±0.0262 | 22.1344±3.3765 |

## V. CONCLUSION

The limited-view PACT is commonly used for most clinical applications, which increases a risk of ill-posed problem. In this work, we investigated the DL-based model-based PA reconstruction and propose the DAV to solve this issue. An iterative adapted gradient scheme, which contains a CNN to extract implicit prior of each iterated result. We believe DAV is a tuning-free iterative method that we do not tune the parameter for each iteration. Furthermore, we hold the data consistency term from the change of CNN. Although this method could show few differences of performance on sane domain data (simulation data in our study), comparing with other learned methods, the results suggest that DAV is more robust with respect to change in the setup. In this work, we only validate 180° view issue for all methods, which could not expose a severe ill-posed condition. In future work, we will validate more different setups, such as the smaller angle less than 90°, and the less number of detectors on different types data.

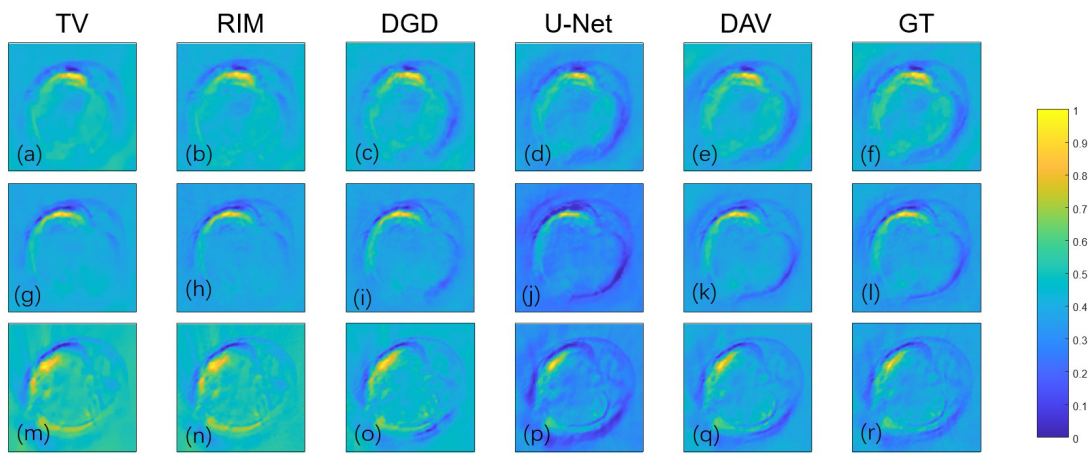

Fig. 4. The samples of *in-vivo* results. (a)-(f) are sample 1; (g)-(l) are sample 2; (m)-(r) are sample 3. (a) (g) (m) total variation with 30 iterations; (b) (h) (n) recurrent inference machines; (c) (i) (o) deep gradient descent; (d) (j) (p) non-iterative U-Net; (e) (k) (q) our method; (f) (l) (r) ground-truth.